# Multi-field approach in mechanics of structural solids

A.A. Vasiliev [a], S.V. Dmitriev [b], A.E. Miroshnichenko [c,*]

[a] *Department of Mathematical Modelling, Tver State University, Sadoviy per. 35, 170002 Tver, Russia*
[b] *Institute for Metals Superplasticity Problems RAS, Khalturina 39, 450001 Ufa, Russia*
[c] *Nonlinear Physics Centre, Research School of Physical Sciences and Engineering,
The Australian National University, Canberra ACT 0200, Australia*

______________________________________________________________________

**Abstract**

We overview the basic concepts, models, and methods related to the multi-field continuum theory of solids with complex structures. The multi-field theory is formulated for structural solids by introducing a macrocell consisting of several primitive cells and, accordingly, by increasing the number of vector fields describing the response of the body to external factors. Using this approach, we obtain several continuum models and explore their essential properties by comparison with the original structural models. Static and dynamical problems as well as the stability problems for structural solids are considered. We demonstrate that the multi-field approach gives a way to obtain families of models that generalize classical ones and are valid not only for long-, but also for short-wavelength deformations of the structural solids. Some examples of application of the multi-field theory and directions for its further development are also discussed.

______________________________________________________________________

## 1. Introduction

A hundred years have passed since the pioneering work by Cosserat E. and Cosserat F. (1909) where they suggested a way for generalization of the elasticity theory. Generalized continuum theories have been extensively developing in various directions since 1950-60s but the discussions on further development of such theories and their applications are not closed until now.

### 1.1. Why continuum modeling?

This question can be answered by analyzing the reasons of success of such continuum theories as the conventional elasticity theory, the reasons why it is useful in modeling of lattice structures, and why the homogenized theories for composites were developed.

______________________________________________________________________

[*] Corresponding author. Tel.: +61-2-6125-9653; fax: +61-2-6125-8588.
*E-mail address:* andrey.miroshnichenko@anu.edu.au (A.E. Miroshnichenko).

Main reasons are the following: continuum models help to define macro-characteristics of structured systems and their expressions through structural parameters; they make it possible to use well-developed mathematical apparatus and methods for developing the theory and to find analytical solutions; in cases when analytical solutions cannot be found, one can use the effective numerical methods and software packages based on artificial discretization with mesh size greater than the original cell size of the body; field theories, having deep history and traditions, represent a well developed set of interrelated theories of continuum mechanics and physics.

However, conventional continuum theories have their limitations that may result in essential errors in the modeling of some specific effects or even in qualitative disagreements with experimental observations. Investigation of such effects within the framework of a field theory demands corrections of the conventional theories and leads to the necessity of generalized continuum models.

*1.2. Generalized continuum mechanics: three possible approaches*

A starting point for the construction of a generalized field theory is the critical analysis and evaluation of the key physical hypotheses and assumptions of the conventional theory. Some of those assumptions can be used by a new theory while others can be weakened or rejected. Such analysis of different generalized theories has been already performed, for example, by Lomakin (1970) and by Rogula (1985).

Let us describe several basic approaches that can be employed for construction of generalized theories descriptive of solids with complex microstructure.

*1.2.1. Intra-cell approach*

Here additional internal degrees of freedom for a unit cell are taken into account. For example, in Cosserat model and in micropolar models, rotational degrees of freedom of structural elements are taken into account in addition to translational degrees of freedom (Cosserat E. and Cosserat F., 1909; Eringen, 1968). Microscopic rotational degrees of freedom can be important for the bodies with elements having finite sizes and for the bodies with a beam-like microstructure. Examples are granular media (Limat, 1988; Suiker et al., 2001; Pasternak and Mühlhaus, 2000, 2002, 2005; Grekova and Herman, 2004), beam lattices (Bažant and Christensen, 1972; Noor, 1988), masonry walls (Casolo, 2004), structured materials (Lakes, 1991; Forest et al., 2001), bones, fabric materials, crystal lattices (Vasiliev et al., 2002, 2005; Pavlov et al., 2006; Ivanova et al., 2007; Potapov et al., 2009); dielectric crystals (Pouget et al., 1986; Askar, 1986; Maugin, 1999), and thin films (Randow et al., 2006) among others.

*1.2.2. Non-local and higher-order gradient models*

Another approach to develop continuum models is taking into account the non-locality (Kröner, 1968; Kunin, 1982; Rogula, 1985; Pasternak and Mühlhaus, 2000, 2002, 2005). Higher order gradient terms for the fields are taken into account in higher-order gradient theories (Aifantis, 1992; Triantafyllidis and Bardenhagen, 1993; Fleck and Hutchinson, 1997, 2001; Peerlings et al., 2001; Bažant and Jirásek, 2002; Askes et al., 2002; Kevrekidis et al., 2002; Aifantis, 2003; Metrikine and Askes, 2006; Andrianov and Awrejcewicz, 2008; Askes et al., 2008; Forest, 2009).

*1.2.3. Macrocell approach*

New class of models can be obtained by considering a macrocell consisting of several primitive translational cells and, accordingly, by increasing the number of vector fields in order to give a better description of the behavior of the original discrete system. This basic idea leads to multi-field models.

The macrocell method and the multi-field theory are the central subjects of this article.

The abovementioned approaches can be used in various combinations allowing one to describe qualitatively different effects in bodies with microstructure. Simplest version of the multi-field theory includes the low-order gradient terms, but a higher-order gradient multi-field theory can also be derived. By increasing the number of fields, the multi-field approach gives a natural way to describe both long- and short-wavelength deformations. One can also apply the multi-field theory to a body with microscopic rotational degrees of freedom thus deriving a micropolar multi-field theory. By using combined models one can describe, for example, short-wavelength distortions in bodies with rotating particles. Furthermore, a combination of these three ideas would result in a micropolar higher-order gradient multi-field theory.

This paper is organized as follows. The basic idea of the multi-field approach is rather simple and it is introduced in Section 2 using a harmonic chain as an example. Generalization of the multi-field theory to 2D case and its application to the media with microscopic rotational degrees of freedom is considered in Section 3. Hierarchy of generalized continuum models describing dynamical properties of the Cosserat lattice with increasing accuracy is obtained. The problem of modeling of short-wavelength static and dynamic localized distortions in frame of the multi-field theory is considered in Sections 4. Section 5 is devoted to the stability problems. Particularly, we develop a multi-field approach to calculate critical loads for a discrete system as well as for a periodic structure with continuum cells. Non-linear problems are addressed in Section 6 where we carry out the multi-field modeling of the short-wavelength $N$-periodic and domain wall structures. Exact $N$-periodic and approximate domain wall analytical solutions are obtained by using multi-field models. Section 7 concludes the paper.

## 2. Dynamics of harmonic chain

In order to demonstrate the basic homogenization techniques mentioned in Section 1.2, we consider a chain of point-wise particles of mass $m$ coupled by the elastic springs with the stiffness constant $c$, see Fig. 1(a).

### 2.1. Discrete model

Transverse displacements of particles, $u_n(t)$, are governed by discrete equations of motion

$$m\ddot{u}_n = c(u_{n+1} - 2u_n + u_{n-1}). \tag{2.1}$$

Dispersion relation for the harmonic waves $u_n(t) = \bar{u} e^{i(\omega t - knh)}$ has the form

$$\omega^2 = 4\frac{c}{m}\sin^2\left(\frac{K}{2}\right), \tag{2.2}$$

where we have denoted $K = kh$.

### 2.2. Conventional elasticity theory

In order to construct a simple continuum model, single function $u(x, t)$ describing the displacements of particles, $u(x,t)|_{x=na} = u_n(t)$, is introduced. Replacing in Eq. (2.1)

$$u_{n\pm 1}(t) \to u(x \pm h, t), \tag{2.3}$$

with the help of Taylor series expansions

$$u(x \pm h, t) = \sum_{r=0}^{N} \frac{(\pm h)^r}{r!} \frac{\partial^r u(x,t)}{\partial x^r}, \tag{2.4}$$

retaining derivatives of up to the second order, $N = 2$, we come to the following continuum equation of motion

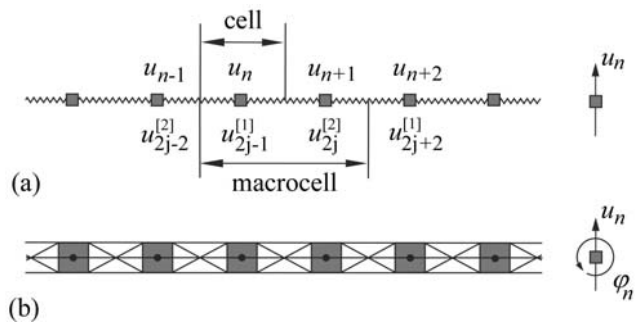

Fig. 1. (a) A one-dimensional chain of particles coupled by elastic springs. (b) A one-dimensional chain of finite size particles with a rotational degree of freedom.

$$\rho u_{tt} = E u_{xx}, \quad (2.5)$$

where $\rho = m/h$ and $E = ch$ are the averaged macroscopic characteristics.

Dispersion relation for harmonic waves, $u(x,t) = \bar{u} e^{i(\omega t - kx)}$, has the form

$$\omega^2 = \frac{c}{m} K^2. \quad (2.6)$$

*2.3. Non-local and higher-order gradient models*

Replacement (2.3) in the discrete equation (2.1) leads to differential in time and difference with respect to space variable non-local equation of motion. After exact Taylor expansion (2.4) with $N = \infty$ we come to the non-local equation which is differential with respect to space variable.

Other forms of non-local continuum models have been developed by Kröner, (1968), Kunin (1982), and Rogula (1985). We will not further discuss this type of continuum models here.

Retaining derivatives of higher then second order in Taylor expansion of displacements leads to the higher-order gradient models. In case $N = 4$, we come to the equation of the higher-order gradient theory for the chain

$$\rho u_{tt} = E\left(u_{xx} + \frac{h^2}{12} u_{xxxx}\right). \quad (2.7)$$

This equation, containing micro-structural parameter $h$, is weakly non-local. Its dispersion relation has the form

$$\omega^2 = \frac{c}{m}\left(K^2 - \frac{1}{12} K^4\right). \quad (2.8)$$

Diffrent variants of higher-order gradient elasticity models have been constructed and discussed, for example, by Aifantis (1992, 2003), Triantafyllidis and Bardenhagen (1993), Askes et al. (2002, 2008). Continuum models derived on the basis of Pade approximations and their implementation in modeling of a spatially discrete system have been discussed by Kevrekidis et al. (2002), Andrianov and Awrejcewicz (2008).

*2.4. Macrocell approach: multi-field models*

The conventional long-wavelength model equation (2.5) and the higher-order gradient model equation (2.7) were obtain by using single primitive translational cell of the chain and single function $u(x,t)$ describing displacements of the particles. We should note that these are some hypotheses accepted in the single-field models; they are not evident and may be rejected.

In order to obtain equations of two-field model, we consider a macrocell that includes two particles and use different notations $u^{[1]}_{2j-1}$ and $u^{[2]}_{2j}$ for the displacements of odd and even particles of the chain. Discrete equations of motion for the particles of a macrocell have the following form

$$m\ddot{u}^{[1]}_{2j-1} = c\left(u^{[2]}_{2j} - 2u^{[1]}_{2j-1} + u^{[2]}_{2j-2}\right),$$
$$m\ddot{u}^{[2]}_{2j} = c\left(u^{[1]}_{2j+1} - 2u^{[2]}_{2j} + u^{[1]}_{2j-1}\right). \quad (2.9)$$

Dispersion relation for the wave solution $u^{[1]}_{2j-1}(t) = \bar{u}^{[1]} e^{i(\omega t - [2j-1]kh)}$, $u^{[2]}_{2j}(t) = \bar{u}^{[2]} e^{i(\omega t - 2jkh)}$ consists of two branches

$$\omega^2 = 4\frac{c}{m} \sin^2\left(\frac{K}{2}\right), \quad (2.10)$$

$$\omega^2 = 4\frac{c}{m} \cos^2\left(\frac{K}{2}\right), \quad (2.11)$$

defined for $0 \leq K \leq \pi/2$. First dispersion relation, Eq. (2.10), coincides with the dispersion relation Eq. (2.2), obtained for the primitive translational cell. Second dispersion relation, Eq. (2.11), can be derived from Eq. (2.2) by replacing $K \to \pi - K$. Hence, the dispersion curve of the discrete model derived for the macrocell can be obtained from Eq. (2.2), derived for the primitive cell, by folding with respect to the line $K = \pi/2$.

In order to obtain two-field model we use two functions $u^{[1]}(x,t)$ and $u^{[2]}(x,t)$ describing displacements of particles in the chain. Assuming that $u^{[1]}(x,t)\big|_{x=(2j-1)h} = u_{2j-1}(t)$, $u^{[2]}(x,t)\big|_{x=2jh} = u_{2j}(t)$ and using Taylor series expansion with derivatives up to the second order we obtain from Eq. (2.9) the equations for the two-field theory

$$mu_{tt}^{[1]} = c\left(2u^{[2]} - 2u^{[1]} + h^2 u_{xx}^{[2]}\right),$$
$$mu_{tt}^{[2]} = c\left(2u^{[1]} - 2u^{[2]} + h^2 u_{xx}^{[1]}\right). \quad (2.12)$$

In terms of new field functions, $u = \tfrac{1}{2}\left(u^{[2]} + u^{[1]}\right)$ and $\tilde{u} = \tfrac{1}{2}\left(u^{[2]} - u^{[1]}\right)$, Eqs. (2.12) become uncoupled,

$$\rho u_{tt} = E u_{xx},$$
$$m\tilde{u}_{tt} = c\left(-4\tilde{u} - h^2 \tilde{u}_{xx}\right). \quad (2.13)$$

The uncoupled form of equations is very useful for analysis and for finding multi-field solutions.

It is interesting to rewrite Eq. (2.12) in the form

$$mu_{tt}^{[1]} = ch^2 u_{xx}^{[1]} + c\left[2\left(u^{[2]} - u^{[1]}\right) + h^2 \left(u^{[2]} - u^{[1]}\right)_{xx}\right],$$
$$mu_{tt}^{[2]} = ch^2 u_{xx}^{[2]} - c\left[2\left(u^{[2]} - u^{[1]}\right) + h^2 \left(u^{[2]} - u^{[1]}\right)_{xx}\right],$$

where we have separated the operators of the conventional elasticity theory, Eq. (2.5), and the additional operator, which describes the interaction of the fields. This representation of the model may be useful for interpretation and further development of the multi-field theory by introducing different hypothesis in the models for basic fields and their interactions.

The dispersion relations of the two-field model, Eq. (2.12), are

$$\omega^2 = \frac{c}{m} K^2, \quad (2.14)$$

$$\omega^2 = \frac{c}{m}\left(4 - K^2\right). \quad (2.15)$$

First dispersion relation, Eq. (2.14), is Taylor series expansion of the dispersion relation for discrete model defined by Eq. (2.2) at the point $K = 0$. Second dispersion relation, Eq. (2.15), is Taylor series expansion in the vicinity of $K = 0$ of the dispersion relation defined by Eq. (2.11). Taking into account the above discussion of the dispersion relations for primitive cell, Eq. (2.2), and macrocell, Eq. (2.11), one can conclude that the dispersion curve defined by Eq. (2.15) for the two-field model in the interval $0 < K < \pi/2$, being folded with respect to the line $K = \pi/2$ on the interval $\pi/2 < K < \pi$, approximates the dispersion curve of the discrete model at the point $K = \pi$.

### 2.5. Intra-cell approach: micropolar model

We have considered two types of generalized continuum models for the chain, Eq. (2.1), shown in Fig. 1(a). Further generalization of the model may be needed in cases when internal degrees of freedom are to be taken into account. For example, for the chain of particles having finite size, shown in Fig. 1(b), a micropolar continuum model can be derived. Such system will be considered in Section 3.

### 2.6. Comparison of the models

Dispersion curves for the discrete and conventional continuum models are shown in Figs. 2(a) and 2(b) by solid and dotted lines, respectively, in dimensionless form, $\bar{\omega} = \omega\sqrt{m/c}$. The dispersion curve for conventional continuum model is tangent line to the dispersion curve of the discrete system at the point $(k, \omega) = (0, 0)$. Thus, the conventional continuum model is applicable for describing long-wavelength waves. It defines group velocity but does not describe dispersion of such waves in the chain. One can see that the difference of the dispersion curves is large for short-wavelength waves.

Dispersion curve for the higher-order gradient model, shown in Fig. 2(a) by dashed line, gives a better approximation for discrete system in the long-wavelength region. It describes dispersion of the waves, but it is still incorrect for the short-wavelength waves.

The dispersion curves for the two-field model are shown in Fig. 2(b) by dashed lines. It coincides with the dispersion curve of the conventional model in the long-wavelength range. Additionally, the two-field model demonstrates a good approximation for the dispersion curve of the discrete system in the short wave region, where both conventional and higher-order gradient single-field models are inappropriate.

*2.7. Three-, four-field, and combined models*

Let us now consider further development of the multi-field theory.

Three and four-field models are obtained by using macrocells consisting of three and four primitive translational cells and, accordingly, three and four field functions are introduced in order to describe the displacements of particles of the chain. By introducing new fields

$$U^{[1]} = u^{[1]} + u^{[2]} + u^{[3]}, \quad U^{[2]} = -u^{[1]} + u^{[2]}, \quad U^{[3]} = -u^{[1]} + u^{[3]},$$

three coupled equations of the three-field model split into the equation of conventional model

$$mU^{[1]}_{tt} = ch^2 U^{[1]}_{xx}$$

and two additional coupled equations,

$$mU^{[2]}_{tt} = -\tfrac{1}{2}ch^2 U^{[2]}_{xx} - ch(U^{[2]}_x - 2U^{[3]}_x) - 3cU^{[2]},$$
$$mU^{[3]}_{tt} = -\tfrac{1}{2}ch^2 U^{[3]}_{xx} + ch(U^{[3]}_x - 2U^{[2]}_x) - 3cU^{[3]}. \tag{2.16}$$

Dispersion curves of the three-field model consist of three pieces (Fig. 3(a)). The first one coincides with the dispersion curve of conventional model in the range $0 \leq K \leq \pi/3$ and approximates the dispersion curve of discrete system in the long-wavelength region. Two other pieces, corresponding to equations (2.16), approximate the dispersion curve of the discrete system for short waves at the point $K = 2\pi/3$.

The four-field model contains four equations and its dispersion curves split into four pieces (Fig. 3(b)). Two of them coincide with the dispersion curves of the two-field model and approximate the dispersion curves of the discrete system for long and short waves at the points $K = 0$ and $K = \pi$ in intervals $0 \leq K \leq \pi/4$ and $3\pi/4 \leq K \leq \pi$, respectively. Two other equations give dispersion curves approximating the dispersion curve of the discrete system for middle wavelength range in the point $K = \pi/2$. Set of the four coupled equations of the four-field model can be separated in four equations, two of which are equations of the two-field model for long and short waves, Eq. (2.12). Two additional equations specify the two-field model for the middle-wavelength range.

It is important to note that basic hypotheses of the approaches discussed in Section 1.2 are independent and can be used in various combinations. For instance, higher-order two-field model can be obtained by using discrete equations for macrocell, Eq. (2.9), retaining derivatives up to fourth order in Taylor series expansions of the displacements. This model contains equation (2.7) of higher-order gradient single-field model and additional higher-order equation, which refines the equation of long-wavelength two-field model for short-wavelength range, Eq. (2.13). Another example is the multi-field higher-order gradient micropolar model for Cosserat lattice, which will be obtained in Section 3, where we consider a model taking into account rotational degrees of freedom for particles.

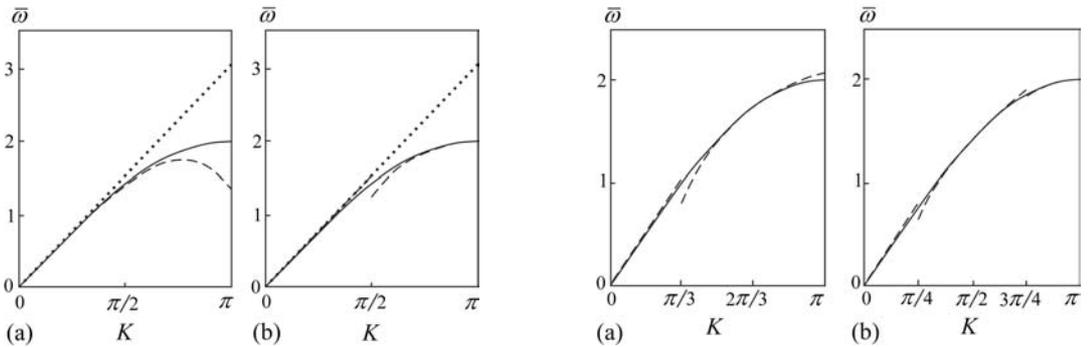

Fig. 2. Dispersion curve of harmonic chain (solid line) and the dispersion curves of the continuum models: conventional elasticity theory (dotted line) and (a) higher-order gradient model and (b) two-field model (dashed lines).

Fig. 3. Dispersion curve of harmonic chain (solid line) and (a) three-field and (b) four-field continuum models (dashed lines).

## 3. Multi-field models for Cosserat solids

### 3.1. Cosserat lattice

We consider a Cosserat lattice, i.e. a lattice where each particle has three degrees of freedom, the displacements, $u_k$, $v_k$, and the rotation $\varphi_k$. The particles are placed at the nodes of a square lattice as shown in Fig. 4(a).

The potential energy associated with the elastic bonds between particles $n$ and $n+1$, in local coordinate system, has the following form

$$E_{pot}^{(n,n+1)} = \tfrac{1}{2} K_n (u_{n+1} - u_n)^2 + \tfrac{1}{2} K_s \left[ v_{n+1} - v_n - \tfrac{1}{2} h (\varphi_{n+1} + \varphi_n) \right]^2 + \tfrac{1}{2} G_r (\varphi_{n+1} - \varphi_n)^2, \quad (3.1)$$

where $h$ is the length parameter, $K_n$ and $K_s$ characterize the stiffness of the bonds in the longitudinal and transverse directions, respectively, and $G_r$ is the stiffness of the local bond preventing rotation of the particle.

Potential energy in the form (3.1) is similar to that considered in the micropolar theory of the elasticity (Eringen, 1968). It is also used in formulation of the discrete models of granular media (Limat, 1988; Pasternak and Mühlhaus, 2000; Suiker et al., 2001), and in the models of micro- and nano-scale thin films (Randow et al., 2006). Potential energy of the beam element (Fig. 4(b)), which is often used for modeling of the beam lattices and for continuum modeling of bodies with a beam-like microstructure (Noor, 1988), is a particular case of the model (3.1). Potential energy in the form Eq. (3.1) may be used for modeling of structural media with finite size particles (Fig. 4(c)). Such models were constructed and micro-structural parameters for the potential energy were found from the experimental data for some crystals by Pavlov et al. (2006) and by Potapov et al. (2009). This model is very useful for better understanding and interpretation of different terms and parameters of the potential energy (3.1).

In order to obtain equations of motion of $k$ th particle, the Lagrangian is constructed. Kinetic energy of $k$ th element has the form $E_{kin}^k = \tfrac{1}{2} M \dot{u}_k^2 + \tfrac{1}{2} M \dot{v}_k^2 + \tfrac{1}{2} J \dot{\varphi}_k^2$.

The equations of motion for $(n,m)$ particle have the form

$$M\ddot{u}_{n,m} = K_n \Delta_{xx} u_{n,m} + K_s \left( \Delta_{yy} u_{n,m} + \tfrac{1}{2} h \Delta_y \varphi_{n,m} \right) + \tfrac{1}{2} K_n^d \left( \Delta u_{n,m} + \Delta_{xy} v_{n,m} \right),$$
$$M\ddot{v}_{n,m} = K_n \Delta_{yy} v_{n,m} + K_s \left( \Delta_{xx} v_{n,m} - \tfrac{1}{2} h \Delta_x \varphi_{n,m} \right) + \tfrac{1}{2} K_n^d \left( \Delta v_{n,m} + \Delta_{xy} u_{n,m} \right), \quad (3.2)$$

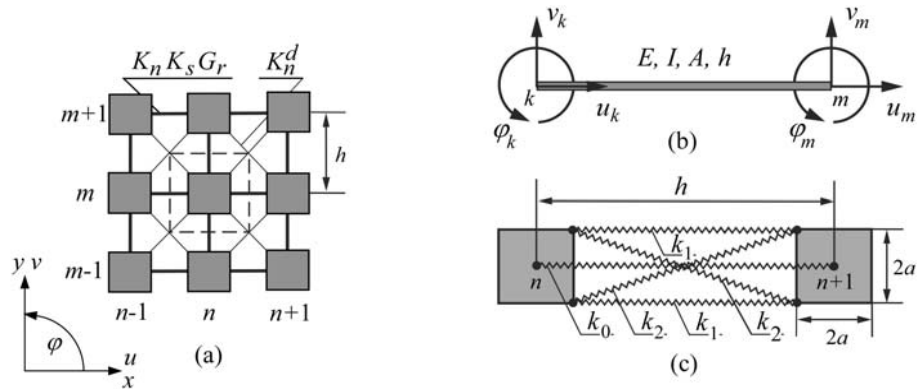

Fig. 4. (a) Square Cosserat lattice: coordinate system and notations. Possible mechanical interpretations of the Cosserat lattice with the potential energy of the form (3.1): (b) beam-like bonding of particles, (c) spring bonding of finite size particles.

$$J\ddot{\varphi}_{n,m} = \left(G_r - \tfrac{1}{4}K_s h^2\right)\left(\Delta_{xx}\varphi_{n,m} + \Delta_{yy}\varphi_{n,m}\right) + \tfrac{1}{2}K_s h\left(\Delta_x v_{n,m} - \Delta_y u_{n,m} - 4h\varphi_{n,m}\right),$$

where the following notations for finite differences are used

$$\Delta_x w_{n,m} = w_{n+1,m} - w_{n-1,m}, \quad \Delta_{xx} w_{n,m} = w_{n+1,m} - 2w_{n,m} + w_{n-1,m},$$
$$\Delta_y w_{n,m} = w_{n,m+1} - w_{n,m-1}, \quad \Delta_{yy} w_{n,m} = w_{n,m+1} - 2w_{n,m} + w_{n,m-1},$$
$$\Delta w_{n,m} = w_{n+1,m+1} + w_{n+1,m-1} - 4w_{n,m} + w_{n-1,m+1} + w_{n-1,m-1},$$
$$\Delta_{xy} w_{n,m} = w_{n+1,m+1} - w_{n+1,m-1} - w_{n-1,m+1} + w_{n-1,m-1}.$$

*3.2. Single-field long-wavelength, higher-order gradient and non-local micropolar models*

In the single-field micropolar model it is assumed that the discrete system can be described by using the vector function $\{u(x,y,t), v(x,y,t), \varphi(x,y,t)\}$, which has the same components as the vector of generalized displacements of the unit cell $\{u_{n,m}(t), v_{n,m}(t), \varphi_{n,m}(t)\}$. It is assumed that the vector-function coincides with vector of displacements at nodes $(nh, mh)$, i.e. $\{u(x,y,t), v(x,y,t), \varphi(x,y,t)\}\big|_{\substack{x=nh\\y=mh}} = \{u_{n,m}(t), v_{n,m}(t), \varphi_{n,m}(t)\}$.

Substituting in the discrete equations $w_{n\pm1,m\pm1}$ with $w(x\pm h, y\pm h)$, and using Taylor series expansions

$$w(x\pm h, y\pm h, t) = \sum_{r=0}^{N_x}\sum_{p=0}^{N_y} \frac{(\pm h)^r}{r!}\frac{(\pm h)^p}{p!}\frac{\partial^{r+p} w(x,y,t)}{\partial x^r \partial y^p} \tag{3.3}$$

one obtains a set of equations which are differential with respect to spatial and temporal variables. They are exact in case $N_x = \infty$, $N_y = \infty$. Taking into account the derivatives up to $N$ th order, for $N > 2$ we come to an approximate higher-order gradient model. Keeping derivatives up to the second order we obtain the long-wavelength single-field equations of micropolar model

$$\rho u_{tt} = \left(K_n + K_n^d\right)u_{xx} + \left(K_s + K_n^d\right)u_{yy} + 2K_n^d v_{xy} + K_s \varphi_y,$$
$$\rho v_{tt} = \left(K_s + K_n^d\right)v_{xx} + \left(K_n + K_n^d\right)v_{yy} + 2K_n^d u_{xy} - K_s \varphi_x, \tag{3.4}$$
$$j\varphi_{tt} = \left(G_r - \tfrac{1}{4}K_s h^2\right)\left(\varphi_{xx} + \varphi_{yy}\right) + K_s\left(v_x - u_y - 2\varphi\right),$$

where $\rho = M/h^2$, $j = J/h^2$.

Single-field model has been obtained in Suiker et al. (2001) where the authors have also found the relation between micro-structural parameters of discrete model and macroscopic parameters by comparing with conventional form of micropolar model. Higher-order gradient micropolar model with derivatives up to fourth order was obtained by Pavlov et al. (2006). Non-local Cosserat model for Cosserat lattice has been derived and studied by Pasternak and Mühlhaus (2000, 2002, 2005). Reduced variant of Cosserat model has been discussed by Grekova and Herman (2004).

*3.3. Hierarchy of multi-field micropolar models*

In order to derive the $N$-field model, we consider a macrocell consisting of $N$ elementary cells (Vasiliev and Miroshnichenko, 2005; Vasiliev et al., 2005, 2008). Although all elements of the lattice within the macrocell are identical (Fig. 4(a)) they are marked with different numbers (examples are shown in Figs. 5(a)-(d)). We use the notations $u_{n,m}^{[s]}(t)$, $v_{n,m}^{[s]}(t)$, $\varphi_{n,m}^{[s]}(t)$ with additional superscript $s = \overline{1,N}$ for the components of the generalized displacements vector for the macrocell. Then, $3N$ discrete equations of motion for the particles of the macrocell, marked by different indices, are constructed. $N$ vector functions $\{u^{[s]}(x,y,t), v^{[s]}(x,y,t), \varphi^{[s]}(x,y,t)\}$, $s = \overline{1,N}$, are introduced in the $N$-field theory in order to describe the displacements and rotations of particles with indices $s = \overline{1,N}$, respectively. By using Taylor series expansions of displacements and rotations in the discrete equations, we come to the equations of the $N$-field theory.

*3.3.1. Two-field micropolar models*

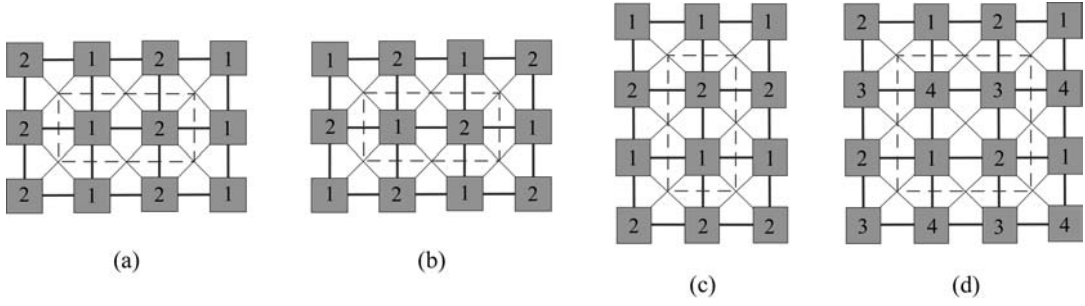

Fig. 5. Macrocells with different choice of particle numbering used in the derivation of (a)-(c) the three different two-field micropolar models and (d) four-field micropolar model.

By using the procedure described above, three types of the two-field models, $N = 2$, corresponding to macrocells presented in Figs. 5(a)-(c) can be obtained. These models will be referred to as "a", "b", and "c" models, respectively. In terms of the new field functions defined as

$$u = \tfrac{1}{2}(u^{[2]} + u^{[1]}), \quad v = \tfrac{1}{2}(v^{[2]} + v^{[1]}), \quad \varphi = \tfrac{1}{2}(\varphi^{[2]} + \varphi^{[1]}),$$
$$\widetilde{u} = \tfrac{1}{2}(u^{[2]} - u^{[1]}), \quad \widetilde{v} = \tfrac{1}{2}(v^{[2]} - v^{[1]}), \quad \widetilde{\varphi} = \tfrac{1}{2}(\varphi^{[2]} - \varphi^{[1]}), \qquad (3.5)$$

the original sets of six equations can be separated into two uncoupled sets. One of them relates the components $u(x,y,t)$, $v(x,y,t)$ and $\varphi(x,y,t)$ and it coincides with the set of equations of the micropolar theory, Eq. (3.4). Therefore, any of three two-field micropolar models, "a", "b", and "c", include the corresponding single-field conventional model and thus, they equally describe the long-wave deformations. The second set of equations for the components $\widetilde{u}(x,y,t)$, $\widetilde{v}(x,y,t)$ and $\widetilde{\varphi}(x,y,t)$ varies with the model. For the model "a", corresponding to the numbering of elements shown in Fig. 5(a), additional set of equations has the form

$$\rho \widetilde{u}_{tt} = -(K_n + K_n^d)\widetilde{u}_{xx} + (K_s - K_n^d)\widetilde{u}_{yy} - 2K_n^d \widetilde{v}_{xy} - 4(K_n + K_n^d)\widetilde{u}/h^2 + K_s \widetilde{\varphi}_y,$$
$$\rho \widetilde{v}_{tt} = -2K_n^d \widetilde{u}_{xy} - (K_s + K_n^d)\widetilde{v}_{xx} + (K_n - K_n^d)\widetilde{v}_{yy} - 4(K_s + K_n^d)\widetilde{v}/h^2 + K_s \widetilde{\varphi}_x, \qquad (3.6)$$
$$j\widetilde{\varphi}_{tt} = (G_r - \tfrac{1}{4}K_s h^2)(-\widetilde{\varphi}_{xx} + \widetilde{\varphi}_{yy}) - K_s(\widetilde{u}_y + \widetilde{v}_x) - 4(G_r + \tfrac{1}{4}K_s h^2)\widetilde{\varphi}/h^2.$$

For the model "b", related to Fig. 5(b), additional set of equations has the form

$$\rho \widetilde{u}_{tt} = (-K_n + K_n^d)\widetilde{u}_{xx} + (-K_s + K_n^d)\widetilde{u}_{yy} + 2K_n^d \widetilde{v}_{xy} - 4(K_n + K_s)\widetilde{u}/h^2 - K_s \widetilde{\varphi}_y,$$
$$\rho \widetilde{v}_{tt} = 2K_n^d \widetilde{u}_{xy} + (-K_s + K_n^d)\widetilde{v}_{xx} + (-K_n + K_n^d)\widetilde{v}_{yy} - 4(K_n + K_s)\widetilde{v}/h^2 + K_s \widetilde{\varphi}_x, \qquad (3.7)$$
$$j\widetilde{\varphi}_{tt} = -(G_r - \tfrac{1}{4}K_s h^2)(\widetilde{\varphi}_{xx} + \widetilde{\varphi}_{yy}) + K_s(\widetilde{u}_y - \widetilde{v}_x) - 8G_r \widetilde{\varphi}/h^2.$$

Finally, for the model "c", related to Fig. 5(c), additional set of equations has the form

$$\rho \widetilde{u}_{tt} = (K_n - K_n^d)\widetilde{u}_{xx} - (K_s + K_n^d)\widetilde{u}_{yy} - 2K_n^d \widetilde{v}_{xy} - 4(K_s + K_n^d)\widetilde{u}/h^2 - K_s \widetilde{\varphi}_y,$$
$$\rho \widetilde{v}_{tt} = -2K_n^d \widetilde{u}_{xy} + (K_s - K_n^d)\widetilde{v}_{xx} - (K_n + K_n^d)\widetilde{v}_{yy} - 4(K_n + K_n^d)\widetilde{v}/h^2 - K_s \widetilde{\varphi}_x, \qquad (3.8)$$
$$j\widetilde{\varphi}_{tt} = (G_r - \tfrac{1}{4}K_s h^2)(\widetilde{\varphi}_{xx} - \widetilde{\varphi}_{yy}) + K_s(\widetilde{u}_y + \widetilde{v}_x) - 4(G_r + \tfrac{1}{4}K_s h^2)\widetilde{\varphi}/h^2.$$

The meaning of the equations (3.6)-(3.8) of the two-field models will be clarified in Section 3.4.

### 3.3.2. Four-field micropolar model

The four-field model corresponds to the macrocell shown in Fig. 5(d). The set of 12 equations for four vector fields of generalized displacements, $u^{[s]}(x,y,t)$, $v^{[s]}(x,y,t)$, $\varphi^{[s]}(x,y,t)$, $s = \overline{1,4}$, can be separated into four sets of equations, (3.4), (3.6), (3.7), and (3.8). Thus, the four-field model includes the long-wave single-field Cosserat model derived in Section 3.2 and all the two-field models derived in Section 3.3.1 by using the macrocells shown in Figs. 5(a)-(c).

Note that the additional sets of equations of the four-field model, Eqs. (3.6)-(3.8), have the same order as the equations of the micropolar elasticity theory, Eq. (3.4). This may help in finding analytical solutions and in the analysis of the dispersion relations based on polynomial approximations (Vasiliev, 1993; Gonella, 2007).

*3.3.3. Higher-order gradient multi-field micropolar models*

Different approaches to generalization of the conventional elasticity theory described in Section 1.2 can be used in various combinations. A combination of the micropolar and the multi-field theories was described in Sections 3.3.1 and 3.3.2. By retaining in the Taylor expansions (3.3) the derivatives up to fourth order, the models formulated in Sections 3.3.1 and 3.3.2 will be transformed to the higher-order gradient multi-field micropolar models. Higher-order gradient two-field micropolar model for the macrocell shown in Fig. 5(b) was derived in Vasiliev et al. (2008).

*3.4. Plane wave solutions. Comparative analysis of the models*

We compare the original discrete model with different continuum approximations by comparing the properties of the corresponding plane wave solutions

$$\begin{bmatrix} u_{n,m}^{[s]}(t) \\ v_{n,m}^{[s]}(t) \\ \varphi_{n,m}^{[s]}(t) \end{bmatrix} = \begin{bmatrix} \bar{u}^{[s]} \\ \bar{v}^{[s]} \\ \bar{\varphi}^{[s]} \end{bmatrix} \exp[i(\omega t - nK_x - mK_y)], \quad \begin{bmatrix} u^{[s]}(x,y,t) \\ v^{[s]}(x,y,t) \\ \varphi^{[s]}(x,y,t) \end{bmatrix} = \begin{bmatrix} \bar{u}^{[s]} \\ \bar{v}^{[s]} \\ \bar{\varphi}^{[s]} \end{bmatrix} \exp[i(\omega t - k_x x - k_y y)], \quad (3.9)$$

where $\bar{u}^{[s]}$, $\bar{v}^{[s]}$, and $\bar{\varphi}^{[s]}$ are the amplitudes, $\omega$ is the frequency, $k_x$, $k_y$ are the wave numbers, and $K_x = k_x h$, $K_y = k_y h$.

*3.4.1. Single-field micropolar models*

The result of the analysis is illustrated in Fig. 6(a). We fix the parameters $K_n$, $h$, $M$ and take the other parameters in the dimensionless form $\bar{K}_s = K_s / K_n = 1/3$, $\bar{K}_n^d = K_n^d / K_n = 0.74$, $\bar{G}_r = G_r / K_n h^2 = 1/3$,

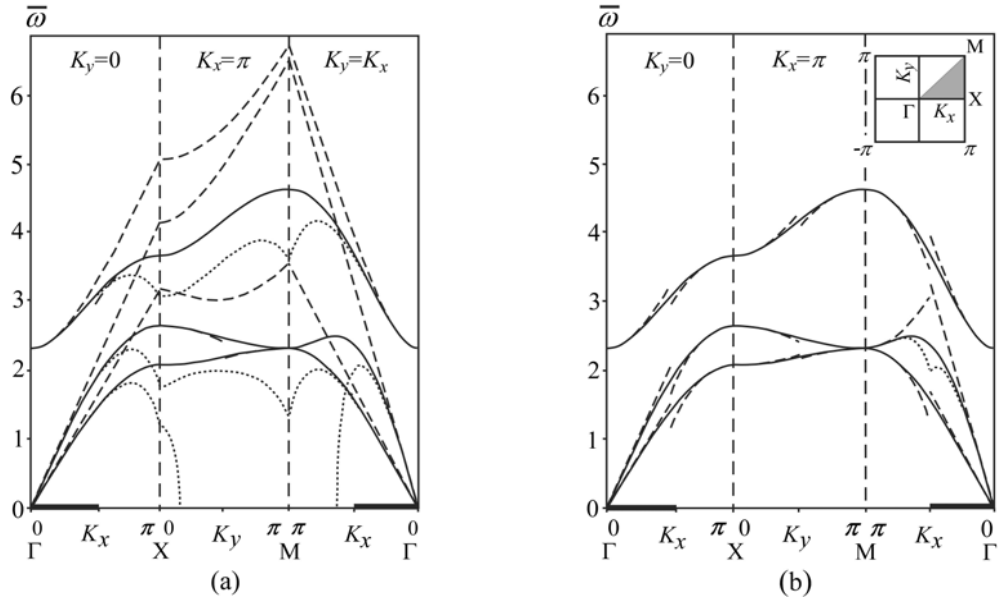

Fig. 6. Sections $K_y = 0$, $K_x = \pi$, $K_x = K_y$ of the dispersion surfaces for the discrete system (solid lines), and the same for (a) the single-field and (b) the four-field models with up to the second and fourth order derivatives, shown by dashed and dotted lines, respectively.

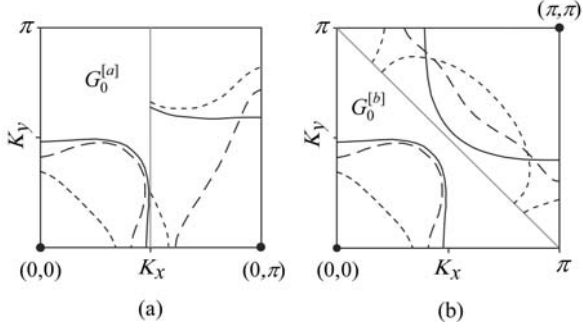 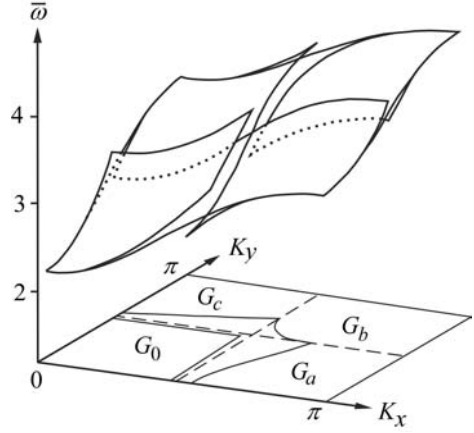

Fig. 7. Regions of the first Brillouin zone near the points $(0,0)$, $(0,\pi)$ and $(0,0)$, $(\pi,\pi)$, where the plane-wave frequencies of the two-field models Eqs. (3.4), (3.6) and Eqs. (3.4), (3.7), corresponding to the macrocells "a" and "b", deviate from the frequencies of the discrete model no more than by 5%.

Fig. 8. The dispersion surface of the micro-rotational waves of the square lattice and its approximation by the four dispersion surfaces of the four-field Cosserat model. Solid line shows the areas near the corners of the first Brillouin zone where the error in the estimation of frequencies of the discrete system is smaller than 10%.

$\bar{I} = I/Mh^2 = 1/8$. Frequency is also presented in dimensionless form $\bar{\omega} = \omega\sqrt{M/K_n}$. Cross sections $K_y = 0$, $K_x = \pi$, and $K_y = K_x$ of the dispersion surfaces for discrete, conventional, and higher-order gradient micropolar models are shown by solid, dashed, and dotted lines, respectively. The dispersion curves of the conventional and higher-order gradient single-field models are tangent to the dispersion curves of the discrete system at the $\Gamma$-point, $(K_x, K_y) = (0, 0)$. The higher-order gradient model improves the accuracy of the approximation of the classical micropolar model for long-wavelength waves. It gives a better description of the wave dispersion. However, for short-wavelength waves, both single-field micropolar models give an essential error.

*3.4.2. Two-field micropolar models*

Six dispersion surfaces of the two-field models shown in Figs. 5(a)-(c) are defined in the regions $G_0^{[a]} = \{0 \leq K_x \leq \pi/2,\ 0 \leq K_y \leq \pi\}$, $G_0^{[b]} = \{0 \leq K_x,\ 0 \leq K_y,\ |K_x + K_y| \leq \pi\}$, $G_0^{[c]} = \{0 \leq K_x \leq \pi,\ 0 \leq K_y \leq \pi/2\}$, respectively, and split into two groups. Three surfaces of the first group are the surfaces of the single-field model, Eq. (3.4), defined in these regions. As it was already mentioned, they approximate the dispersion surfaces of the discrete system for the long waves near the point $(0, 0)$ of the first Brillouin zone. The second group of surfaces corresponds to the additional set of equations (3.6), (3.7) or (3.8). These surfaces being reflected on the areas additional to $G_0^{[a]}$, $G_0^{[b]}$, and $G_0^{[c]}$ in the first Brillouin zone with respect to the planes $K_x = \pi/2$, $K_x + K_y = \pi$, and $K_y = \pi/2$ approximate dispersion surfaces of discrete model for short waves near the points $(\pi, 0)$, $(\pi, \pi)$, and $(0, \pi)$, respectively. Figures 7(a) and 7(b) show the regions near the points $(0,0)$, $(0,\pi)$ and $(0,0)$, $(\pi,\pi)$ of the first Brillouin zone, where the relative error $\left|(\omega_s^{cont.} - \omega_s^{discr.})/\omega_s^{discr.}\right|$, $s = \overline{1,3}$, in the approximation of the spectrum of the discrete system by using "a" and "b" two-field models is smaller than 5%.

Thus, the two-field models include the single-field model that reproduces the dynamical properties of discrete system for long waves. The additional equations (3.6), (3.7), and (3.8) of the two-field models "a", "b", and "c" improve the single-field micropolar model in the short-wavelength range near the points of the first Brillouin zone $(\pi, 0)$, $(\pi, \pi)$, and $(0, \pi)$, respectively (Vasiliev and Miroshnichenko, 2005).

*3.4.3. Four-field micropolar model*

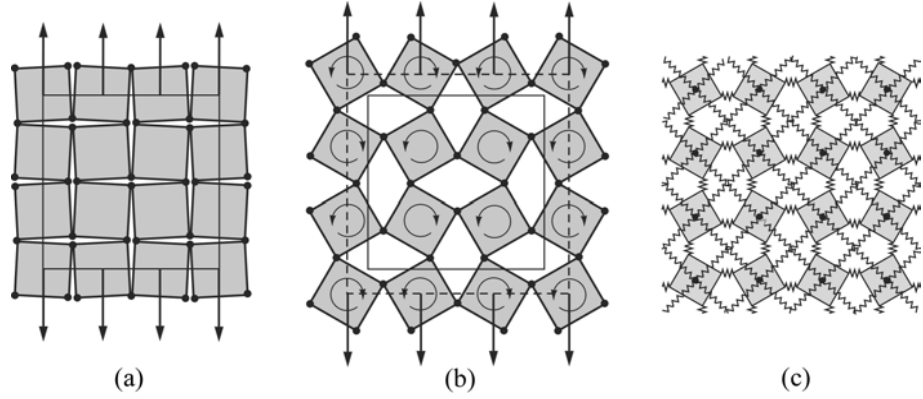

Fig. 9. Auxetic behavior of the lattice with rotating particles: (a) initial configuration and (b) configuration after stretching. The area shown by solid line in (a) transforms to the area shown by dashed line in (b). Material demonstrates the unusual auxetic property (negative Poisson ratio), i.e., extends in lateral direction being uniaxially stretched. (c) Structural model of crystal used in Vasiliev et al. (2002) to investigate the auxetic behavior.

Dispersion surfaces of the four-field model consist of four groups defined in the region $G_0$ of the first Brillouin zone (Fig. 8). The first group includes the surfaces of single-field micropolar model, Eq. (3.4). Three other groups are the surfaces defined in the region $G_0$ by the additional equations (3.6), (3.7), and (3.8). These surfaces for the four-field model, being reflected from the region $G_0$ with respect to the planes $K_x = \pi/2$, $K_x + K_y = \pi$, and $K_y = \pi/2$, on the regions $G_a$, $G_b$, and $G_c$, respectively, approximate the dispersion surfaces of the discrete model for the short waves near the points $(\pi, 0)$, $(\pi, \pi)$, and $(0, \pi)$, respectively.

The dispersion surface for micro-rotational waves obtained by using discrete model is shown in Fig. 8 by solid line for the region $0 \leq K_x \leq \pi$, $0 \leq K_y \leq \pi$. Four dispersion surfaces of the four-field model are shown for the regions $G_0$, $G_a$, $G_b$, and $G_c$. These surfaces are tangent to the dispersion surface of the discrete model at the points $(0,0)$, $(\pi,0)$, $(\pi,\pi)$, and $(0,\pi)$, respectively. The area of the first Brillouin zone near the points $(0,0)$, $(\pi,0)$, $(\pi,\pi)$, and $(0,\pi)$, where the relative error $\left|\left(\omega^{cont.} - \omega^{discr.}\right)/\omega^{discr.}\right|$ of the four-field model is smaller than 10% is shown by solid line.

The sections $K_y = 0$, $K_x = K_y$, $K_x = \pi$ of the dispersion surfaces for discrete, four-field long-wavelength and higher-order gradient micropolar models are shown in Fig. 6(b) by solid, dashed, and dotted lines, respectively. For comparison, the same curves obtained by using discrete, single-field long-wavelength and higher-order gradient micropolar models are shown in Fig. 6(a). The four-field model coincides with the single-field model in the region $G_0$ within the interval shown by thick line on $K_x$ axis for long waves (near $\Gamma$-point), where the single-field model demonstrates a good approximation. The four-field model gives also a good approximation for short waves near the points $X = (\pi, 0)$, $M = (\pi, \pi)$, and $(0, \pi)$.

Before closing this Section, we note that the methods described above have been used for modeling Cosserat solids consisting of finite size particles (see Fig. 9). Such solids can demonstrate auxetic property (negative Poisson ratio). A micropolar continuum model with this property was derived in Vasiliev et al. (2002) and Dmitriev et al. (2005), where the relation between macro- and micro-structural parameters was establishes. It was shown that the rotational degrees of freedom of finite size particles are essential for explaining the negative Poisson ratio exhibited by the model in a certain range of micro-structural parameters. Two- and four-field models taking into account short-wave rotations coupled to the long-wave deformations (Fig. 9(b)) were derived by Vasiliev et al. (2005).

## 4. Multi-field modeling of short-wave localized distortions

Typically, discrete and continuum models are compared by looking at properties of the plane wave solutions of the form (3.9) in both models. It may also be important to check whether a continuum model can describe static and dynamic localized distortions supported by a discrete system, and if yes, what is the accuracy of the description. The localized distortions can appear in the form of localized dynamical excitations, static deformations near surfaces, defects, and concentrated forces in structural solids.

*4.1. One-dimensional localized short-wave static and dynamic solutions in Cosserat lattice*

Here we would like to demonstrate that the Cosserat lattice (Fig. 4(a)) supports such localized distortions that cannot be described either by conventional elasticity, or by the single-field higher-order gradient micropolar models but they can be described by the two-field model discussed in Section 3.3.1.

*4.1.1. One-dimensional models*

The one-dimensional tension-compression of a lattice placed between two rigid slabs (see Fig. 10) is considered under the assumption that the displacements and rotations of particles do not change along the diagonal, i.e., $u_{n,m}$, $v_{n,m}$, and $\varphi_{n,m}$ are constant for elements $(n,m)$ with $n+m = const$. Discrete equations of motion, Eq. (3.2), for displacements $U_m$ in the coordinate system $O\xi\eta$ have the form

$$M\ddot{U}_m = (K_n + K_s)(U_{m-1} - 2U_m + U_{m+1}) + K_n^d(U_{m-2} - 2U_m + U_{m+2}). \quad (4.1)$$

Higher-order gradient one-dimensional single-field model, Eq. (3.4), has the form

$$MU_{tt} = (K_n + K_s + 4K_n^d)H^2 U_{\xi\xi} + \frac{1}{12}(K_n + K_s + 16K_n^d)H^4 U_{\xi\xi\xi\xi}, \quad (4.2)$$

where $H = \sqrt{2}h/2$ is distance between layers.

Similarly, changing the variables and considering one-dimensional displacements, additional equation (3.7) of the two-field model (model "b" in Section 3.3.1) gives the second equation of the one-dimensional model

$$M\widetilde{U}_{tt} = -4(K_n + K_s)\widetilde{U} - (K_n + K_s - 4K_n^d)H^2 \widetilde{U}_{\xi\xi} - \frac{1}{12}(K_n + K_s - 16K_n^d)H^4 \widetilde{U}_{\xi\xi\xi\xi}. \quad (4.3)$$

Equations (4.2) and (4.3) of the two-field model can be derived directly from Eq. (4.1) by using general method.

*4.1.2. Comparison of the localized dynamic solutions in different models*

It is interesting to compare the discrete, the single- and the two-field models with respect to the solutions of the form $U_m(t) = \overline{U}e^{i\omega t - Km}$ and $U(\xi,t) = \overline{U}e^{i\omega t - K\xi/H}$ with complex $K = K_{Re} + iK_{Im}$, i.e. in the case when $K_{Re}$ is not equal to zero.

The dispersion curves of the discrete model have a branch of harmonic waves in the plane $K_{Re} = 0$, a branch of the short wavelength localized solutions in the plane $K_{Im} = \pi$, and a higher frequency branch defined for complex values of $K$ with $K_{Re} \neq 0$, $K_{Im} \neq 0$ which starts from the maximal point of the harmonic or short-wavelength branch. As it was described in Section 3.4, harmonic branches of the long-wave and the higher-order gradient single-field models approximate harmonic branch of the discrete model for long waves but they fail to describe the short-wavelength harmonic waves near the point $K_{Im} = \pi$. Single-field models with derivatives up to the second and even fourth order do not give the short wavelength branch of the localized solutions in the plane $K_{Im} = \pi$. On the other hand, the two-field model demonstrates a good approximation for the branches of harmonic waves in the plane $K_{Re} = 0$ for long and short waves near the points $K_{Im} = 0$ and $K_{Im} = \pi$. Moreover, the two-field model has the branch of short-wavelength localized solutions in the plane $K_{Im} = \pi$, which approximates the branch of the discrete model near the point $(K_{Im}, K_{Re}) = (\pi, 0)$. The higher-order single-field and two-field models give the higher frequency branch for complex values, $K_{Re} \neq 0$, $K_{Im} \neq 0$, but inaccuracy demonstrated by the single-field model is relatively large because this branch for discrete system starts in the short-wave region.

*4.1.3. Static solutions*

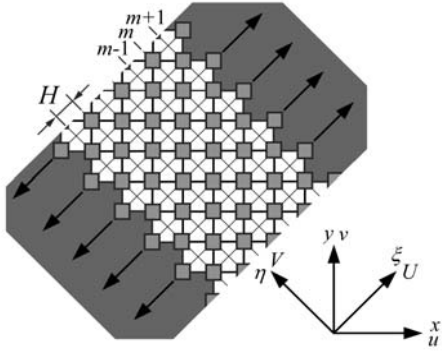 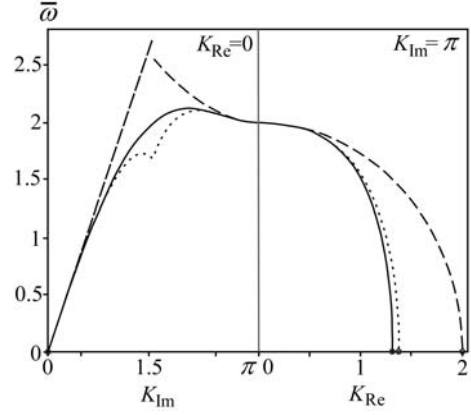

Fig. 10. One-dimensional tension of the lattice layer between rigid slabs.

Fig. 11. Dispersion curves for harmonic, $K_{Re}=0$, and localized short-wave solutions, $K_{Im}=\pi$, obtained for the discrete system and for the two-field models with derivatives up to second and fourth orders, shown by solid, dashed and dotted lines, respectively. Points of intersection of dispersion curves with $\omega=0$ plane define static solutions.

Static solutions of the models are determined by the values $(K_{Im},K_{Re})$ of intersection of dispersion curves with $\omega=0$ plane.

Characteristic equation of the discrete equation (4.1) in the static problem has the roots $(K_{Im},K_{Re})=(0,0)$ of the second order and the roots $(K_{Im},K_{Re})=(\pi,\pm\lambda)$, where $\lambda$ is the solution of the equation $1+\cosh\lambda+\gamma(1-\cosh 2\lambda)=0$, $\gamma=K_n^d/(K_n+K_s)$. General static solution of Eq. (4.1) has the form

$$U_m = C_0 + mC_1 + (-1)^m e^{\lambda m} C_2 + (-1)^m e^{-\lambda m} C_3. \qquad (4.4)$$

The long-wave, single-field model defined by Eq. (4.2) with the derivatives up to the second order gives the root $(K_{Im},K_{Re})=(0,0)$ of the second order, which gives only linear part of the static solution. The single-field model with derivatives up to the fourth order, Eq. (4.2), gives four roots, two of which $(K_{Im},K_{Re})=(0,0)$ give the linear part of the solution, however the two other roots do not give the localized rapidly varying part of the solution because the corresponding branch of the higher-order gradient single-field model belongs to the plane $K_{Re}=0$ and, as it was already mentioned, there are no branches of this model in the plane $K_{Im}=\pi$.

By using Eq. (3.5) we find $U^{[s]}(\xi)=U(\xi)+(-1)^s\widetilde{U}(\xi)$, where $U(\xi)$ and $\widetilde{U}(\xi)$ are solutions to equations (4.2) and (4.3), respectively. The two-field model with up to the second order derivatives gives the following static solution:

$$U^{[s]}(\xi) = c_0 + c_1\xi/H + (-1)^s e^{\Lambda\xi/H} c_2 + (-1)^s e^{-\Lambda\xi/H} c_3, \quad s=1,2. \qquad (4.5)$$

where $\Lambda$ is the root of the equation $4+(1-4\gamma)\Lambda^2=0$.

Solution (4.5) to the two-field model is qualitatively similar to that obtained for the discrete model, Eq. (4.4). Quantitative comparison of the two models stems from the fact that the equation for the parameter $\Lambda$ of the static solution to the two-field model is nothing but the Taylor series expansion of the equation for corresponding parameter $\lambda$ of the discrete solution, $2(1+\cosh\lambda)+2\gamma(1-\cosh 2\lambda)=4+(1-4\gamma)\lambda^2+O(\lambda^4)$. The parameters $\Lambda$ and $\lambda$ are closer to each other in the case of weak localization, $\lambda\approx 0$.

Figure 11 illustrates the results described above. Dispersion curves for harmonic (plane $K_{Re}=\pi$) and localized short-wave solutions (plane $K_{Im}=\pi$) obtained by using the discrete and the two-field models with derivatives up to second and fourth orders are shown by solid, dashed and dotted lines, respectively. Analysis for the harmonic waves was made in Section 3.4. The plane $K_{Re} = 0$ in Fig. 11 corresponds to the plane $K_x = K_y$ in Fig. 6(b). The branch of the two-field model approximates in the plane $K_{Im} = \pi$ the branch for the localized solutions of the discrete model at the point $(K_{Im}, K_{Re}) = (\pi, 0)$. The points of intersections of the dispersion curves with the plane $\omega = 0$ define static solutions (shown by circles). We should note here that although the higher-order gradient two-field model is more accurate, it is also more complicated. Thus, the model with the second order derivatives may be optimal for applications.

*4.2. Short-wavelength localized static distortions near boundaries, defects, and concentrated forces in beam-like bodies*

Micropolar type continuum models were developed as effective methods for analysis of the beam-like and plate-like structures consisting of large number of periodically repeated elements (see a review by Noor, 1988). Such systems can also be considered as simple structural models for thin films or interfaces. Some methods and models for such bodies were considered in Gonella (2007) and Vasiliev (1993, 1996) by introducing multi-cells in dynamic analysis and multi-field approximations. In this Section we will give an example of application of the multi-field approach to the structures of this type.

Let us consider the truss system shown in Fig. 12(a). Lateral tension load $f$ is applied to each node and we take into account the transverse displacements and neglecting the longitudinal ones. Suppose that the element with the number $n = 0$ is broken (Fig. 12(b)).

Static equations have the form

$$G_1 \bar{b}^2 (u_{n+1} + 2u_n + u_{n-1}) + 2G_2 u_n = f, \quad n \neq 0, \tag{4.6}$$

$$G_1 \bar{b}^2 (u_1 + 2u_0 + u_{-1}) = f, \quad n = 0, \tag{4.7}$$

where $\bar{b} = b/r$ with $r = \sqrt{a^2 + b^2}$, $G_1 = E_1 A_1 / r$, $G_2 = E_2 A_2 / b$, with $E_1$ and $E_2$ being stiffness coefficients, $A_1$ and $A_2$ being cross areas, and $r$, $b$ are the lengths of the beam elements.

Symmetric static solution, $u_{-1} = u_1$, to the equations (4.6) and (4.7) can be presented as a sum of slowly varying solution $\bar{u} = f / (4G_1 \bar{b}^2 + 2G_2)$, which can be obtained as constant solution $u_n = \bar{u}$ to the equation (4.6), and the rapidly varying part localized near the defect

$$u_n = \bar{u} + (-1)^n \frac{c}{1 - e^{-\lambda}} \bar{u} e^{-\lambda n}, \quad n \geq 0. \tag{4.8}$$

Here parameter $\lambda > 0$ is the root of the equation $\cosh(\lambda) = 1 + c$, where $c = G_2 / G_1 \bar{b}^2$. This parameter describes the degree of localization of the second part of the solution.

The single-field equation, $(4G_1 \bar{b}^2 + 2G_2) u(x) = f$, which is the long-wave continuum analog of Eq. (4.6), gives only the first part of the solution, $u(x) = \bar{u}$, but it does not produce the rapidly varying part.

The two-field model can be obtained considering the macrocell consisting of two primitive periodic cells of the structure, introducing two functions $u^{[1]}(x)$ and $u^{[2]}(x)$ such that $u^{[1]}((2m-1)a) = u_{2m-1}$, $u^{[2]}(2ma) = u_{2m}$ and using the Taylor series expansions with derivatives up to the second order:

$$2(G_1 \bar{b}^2 + G_2) u^{[1]} + G_1 \bar{b}^2 (2u^{[2]} + a^2 u_{xx}^{[2]}) = f,$$
$$G_1 \bar{b}^2 (2u^{[1]} + a^2 u_{xx}^{[1]}) + 2(G_1 \bar{b}^2 + G_2) u^{[2]} = f. \tag{4.9}$$

Equation (4.7), which is symmetric with respect to $x = 0$, gives the following boundary condition:

$$G_1 \bar{b}^2 [2u^{[1]}(0) + 2a u_x^{[1]}(0) + 2u^{[2]}(0)] = f. \tag{4.10}$$

The solution of the problem (4.9), (4.10) looks as follows:

$$u^{[s]}(x) = \bar{u} + (-1)^s \frac{c}{1 - (1 - \mu)} \bar{u} e^{-\mu \frac{x}{a}}, \quad \mu = \sqrt{2c}, \quad s = 1, 2, \quad x \geq 0. \tag{4.11}$$

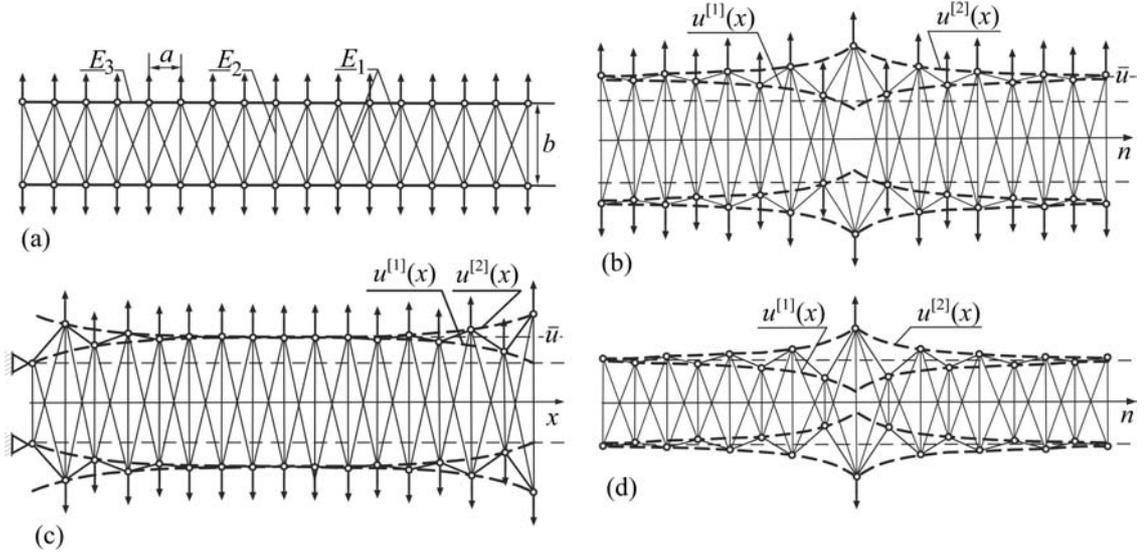

Fig. 12. (a) Periodic structure consisting of pin-jointed beams under transverse load. Two-field approximation of slowly and rapidly varying displacements near (b) defect, (d) boundaries, and (c) concentrated force.

Far from the defect, i.e. for $|n| \gg 1$, the discrete (4.8) and the two-field (4.11) solutions coincide. Near the defect both solutions (4.8) and (4.11) vary rapidly. Equation $\mu^2 - 2c = 0$ for the coefficient $\mu$ defines the decreasing rate of the solution to the two-field model. This equation is a truncated Taylor series expansion of the equation $2[\cosh(\lambda)-1] - 2c = \lambda^2 - 2c + o(\lambda^4)$ for similar coefficient $\lambda$ in the discrete solution. Figure 12(b) illustrates describing of slowly and rapidly varying displacements in discrete system by using two slowly varying functions in two-field model.

Rapidly varying distortions, similar to that appearing near a defect, can also take place near a boundary or near a concentrated force as illustrated by Figs. 12(c) and 12(d). One can see that both slowly and rapidly varying displacements may be accurately described by using two slowly varying functions. The two-field model can be effectively used in these cases too.

## 5. Stability problems

Application of the multi-field theory to the stability problems will be demonstrated for the two examples: (i) stability of the axially loaded chain of hinged rods on elastic supports and (ii) stability of the cylindrical shell periodically stiffened by frames under hydrostatic pressure.

*5.1. Discrete system*

*5.1.1. Discrete model*

Buckling of pin-ended system consisting of $N+1$ rigid rods connected by elastic hinges on elastic supports under axial load (Fig. 13) is a classical problem of stability theory (Timoshenko, 1936). One is looking for the minimal value of the axial load $p$, for which the boundary value problem for the discrete stability equation

$$ch\frac{w_{n+2} - 4w_{n+1} + 6w_n - 4w_{n-1} + w_{n-2}}{h^4} + p\frac{w_{n+1} - 2w_n + w_{n-1}}{h^2} + \frac{k}{h}w_n = 0, \quad n = \overline{1, N}, \qquad (5.1)$$

$$w_0 = 0, \quad w_{-1} - 2w_0 + w_1 = 0, \quad w_{N+1} = 0, \quad w_{N+2} - 2w_{N+1} + w_N = 0,$$

has non-trivial solution. Here $c$ and $k$ are the stiffness constants of hinges and supports, respectively, $h$ is the length of the rods, and $w_n$ are the transverse displacements of hinges (Fig. 13).

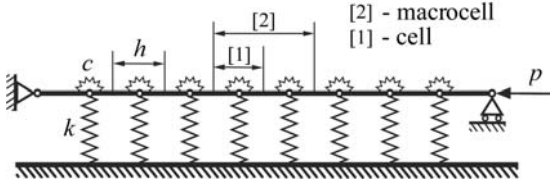
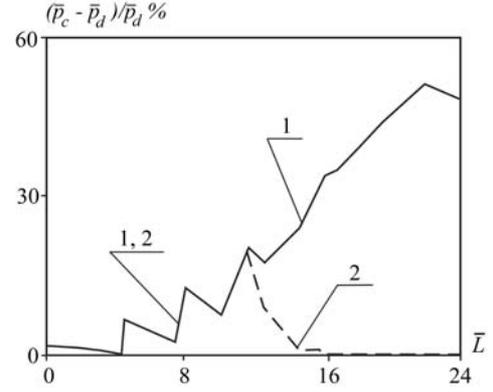

Fig. 13. Elastically supported multi-link system loaded by the axial force *p*.

Fig. 14. Relative error $|(\bar{p}_c - \bar{p}_d)/\bar{p}_d|$ % in case when dimensionless critical loads for the discrete system, $\bar{p}_d$, are obtained by using the single-field (solid line, number "1") and the two-field (dashed line, number "2") models, which give the critical value $\bar{p}_c$.

*5.1.2. Single-field model*

By using Taylor series expansions of displacements in Eq. (5.1) and taking into account derivatives up to fourth order in the stability equation and up to second order in the boundary conditions, we obtain the following stability equation and boundary conditions of the single-field model:

$$ch\, w^{IV} + p\left(w'' + \frac{h^2}{12}w^{IV}\right) + \frac{k}{h}w = 0, \quad 0<x<L, \tag{5.2}$$

$$w(0)=0, \quad w''(0)=0, \quad w(L)=0, \quad w''(L)=0,$$

where $L = (N+1)h$ is the system length. This is the well-known stability problem for a pin-ended beam on the elastic foundation. By obtaining this continuum model, one finds the effective bending rigidity, $ch$, and the effective rigidity of elastic foundation, $k/h$. Let us note that the term $h^2 w^{IV} p/12$ and, hence, parameter $h$ is not taken into account by the classical equation, i.e. the classical model is local.

*5.1.3. Two-field stability model*

The single-field model was derived by using single primitive translational cell of the structure and single function for displacements of hinges describing the buckling mode. In order to obtain the two-field model we consider a macrocell with two primitive translational cells that includes two hinges. Displacements of even and odd hinges are denoted as $u_n$ and $v_n$, respectively. We rewrite Eq. (5.1) for the hinges of a macrocell in the form

$$\frac{ch}{h^4}(u_{n-2} - 4v_{n-1} + 6u_n - 4v_{n+1} + u_{n+2}) + \frac{p}{h^2}(v_{n-1} - 2u_n + v_{n+1}) + \frac{k}{h}u_n = 0,$$

$$\frac{ch}{h^4}(v_{n-1} - 4u_n + 6v_{n+1} - 4u_{n+2} + v_{n+3}) + \frac{p}{h^2}(u_n - 2v_{n+1} + u_{n+2}) + \frac{k}{h}v_{n+1} = 0. \tag{5.3}$$

The two-field model uses two field functions, $u(x)$ and $v(x)$, to describe displacements of hinges. By using Taylor series expansions of displacements taking into account derivatives up to fourth order we come to the set of coupled equations of the two-field model (Vasiliev, 1993). In terms of the new functions, $U = \frac{1}{2}(u+v)$, $V = \frac{1}{2}(u-v)$, this set of equations uncouples:

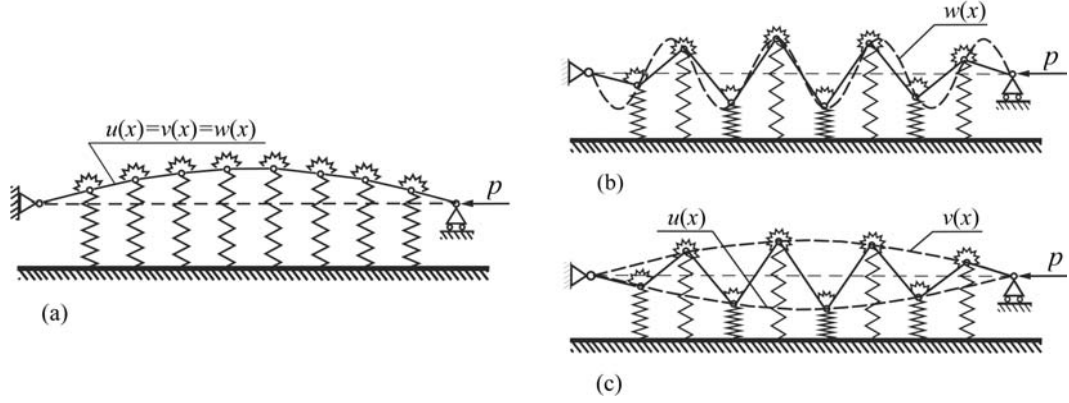

Fig. 15. (a) Approximation of the long-wave buckling mode of the discrete system by slowly varying displacement function of the single-field model and by two functions (they coincide) of the two-field model. (b) The short-wave buckling mode varies rapidly and the single-field model fails in this case. (c) The short-wave buckling mode is accurately described by two slowly varying functions of the two-field model.

$$ch\, U^{IV} + p\left(U'' + \frac{h^2}{12}U^{IV}\right) + \frac{k}{h}U = 0, \tag{5.4}$$

$$4ch\left(\frac{5}{12}V^{IV} + \frac{2}{h^2}V'' + \frac{4}{h^4}V\right) - p\left(\frac{4}{h^2}V + V'' + \frac{h^2}{12}V^{IV}\right) + \frac{k}{h}V = 0. \tag{5.5}$$

*5.1.4. Comparison of the model*

The relative error of the single-field model in estimation of the dimensionless critical loads $\bar{p} = p/\sqrt{kc}$ of discrete system consisting of nine rods, $N = 8$, is shown in Fig. 14 by solid line as the function of dimensionless parameter $\bar{L} = L\sqrt[4]{(k/h)/ch}$. One can see that the single-field model gives a good accuracy for small values of $\bar{L}$, when long-wave buckling modes take place and gives a considerable error for large values of $\bar{L}$, when the buckling modes are of short-wavelength type. We should note that the short-wavelength buckling modes take place very often in practice in the case of sufficiently rigid supports.

The first equation (5.4) of the two-field model coincides with the stability equation of the single-field theory, Eq. (5.2). Hence, the two-field model includes the single-field model and thus, it describes the long-wave buckling of the system with the same accuracy. The second equation (5.5) improves the conventional continuum model, equation (5.2), for the short-wavelength modes. The relative error of the two-field model in the estimation of the critical loads of the discrete system is shown in Fig. 14 by the dashed line for different values of $\bar{L}$. The two-field model is accurate not only for the long-wave buckling modes (small $\bar{L}$), but also for the short-wave modes (large $\bar{L}$). Let us note that the two-field model improves the single-field model in the range of parameters where the latter one gives the greatest error.

Figures 15(a)-(c) explain the results for the two-field model. The long-wave buckling modes can be accurately described by using single slowly varying function (Fig. 15(a)) and the single-field model with the lowest order derivatives is sufficiently accurate in this case. On the other hand, the short-wave modes vary rapidly (Fig. 15(b)) and the model defined by Eq. (5.2) with the lowest order derivatives gives a very poor accuracy. Figures 15(a) and 15(c) show that both long-wave and short-wave buckling modes can be accurately described by using two slowly varying functions and hence, the two-field model demonstrates a good accuracy in both cases.

*5.2. Continuum system*

Here we demonstrate how to obtain a multi-field model in the case when primitive translational cell is a continuum system. As an example, we consider a stability problem for cylindrical shell periodically stiffened by elastic frames and loaded by uniform external pressure $p$ (Fig. 16). We use a structural model considered in the work (Alfutov, 2000). The stability equation for the shell between neighboring frames, $ma \leq x \leq (m+1)a$, has the form

$$L\Phi_m = 0, \quad L = B_x \frac{\partial^4}{\partial x^4} + \frac{D_\varphi}{R^6} \frac{\partial^4}{\partial \varphi^4}\left(\frac{\partial^4}{\partial \varphi^4} + 2\frac{\partial^2}{\partial \varphi^2} + 1\right) + \frac{p}{R^3}\frac{\partial^4}{\partial \varphi^4}\left(\frac{\partial^2}{\partial \varphi^2} + 1\right), \quad (5.6)$$

where $\Phi_m(x,\varphi)$ is the displacement function. The equation for circumferential displacement $V_m(\varphi)$ of the of the axial line of $m$-th frame is

$$EJ \frac{\partial^2}{\partial \varphi^2}\left(\frac{\partial^4 V_m}{\partial \varphi^4} + 2\frac{\partial^2 V_m}{\partial \varphi^2} + V_m\right) = R^4\left(S_{m-1}(a) - S_m(0)\right), \quad (5.7)$$

where $S_m$ is shear force, $\frac{\partial S_m}{\partial \varphi} = -B_x R \frac{\partial^3 \Phi_m}{\partial x^3}$. There are four compatibility conditions for each frame.

We will eliminate the internal degrees of freedom describing the shell spans and obtain the discrete systems describing the behavior of the frames. The single-field and the multi-field models can be obtained for these discrete equations by using the technique described in above.

We look for solutions to Eq. (5.6) of the form $\Phi_m(x,\varphi) = X_m(x)e^{in\varphi}$. Substituting this into Eq. (5.6) we come to an ordinary differential equation of fourth order for the function $X_m(x)$ for the shell section between $m$th and ($m+1$)th frames. Solution of this equation includes four integration constants. By using the four compatibility conditions the integration constants can be eliminated and the following finite difference equations for $\Phi_m^0 = X_m(0)e^{in\varphi}$ can be obtained:

$$\Phi_{m+2}^0 + 2\Phi_m^0 + \Phi_{m-2}^0 - f(\bar{\lambda})\left(\Phi_{m+1}^0 + \Phi_{m-1}^0\right) + g(\bar{\lambda})\Phi_m^0 = 0,$$

where

$$f(\bar{\lambda}) = 2(\cos\bar{\lambda} + \cosh\bar{\lambda}) + \frac{c}{2\bar{\lambda}^3}(\sin\bar{\lambda} - \sinh\bar{\lambda}), \quad g(\bar{\lambda}) = 4\cosh\bar{\lambda}\cos\bar{\lambda} + \frac{c}{\bar{\lambda}^3}(\cosh\bar{\lambda}\sin\bar{\lambda} - \sinh\bar{\lambda}\cos\bar{\lambda}),$$

$$\bar{\lambda} = \lambda\bar{a}, \quad \bar{a} = \frac{a}{R}, \quad \lambda^4 = \frac{R}{B_x}n^4(n^2-1)\left[p - \frac{D_\varphi}{R^3}(n^2-1)\right], \quad c = \frac{EJ\bar{a}^3}{B_x R^3}n^4(n^2-1)^2.$$

By using Taylor series expansion of functions $g(\bar{\lambda})$ and $f(\bar{\lambda})$,

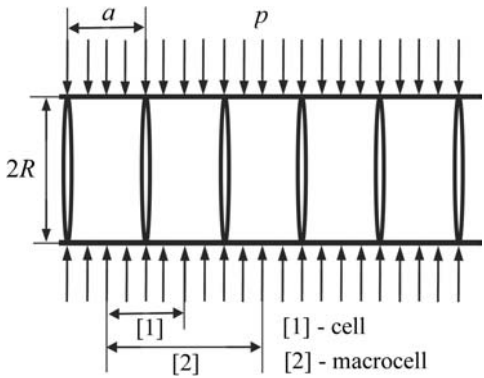 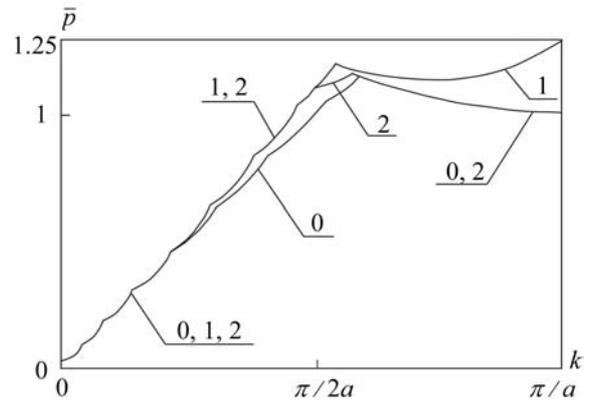

Fig. 16. Cylindrical shell periodically stiffened with elastic frames and loaded with uniform external pressure *p*.

Fig. 17. Dimensionless critical pressure as the function of wave number of the buckling mode calculated by using the structural inhomogeneous, the homogeneous single-field, and the homogeneous two-field models, shown by curves "0", "1", and "2", respectively.

$$f(\bar{\lambda}) = \sum_{r=0}^{\infty} \left[ \frac{4}{(4r)!} + \frac{-1}{(4r+3)!} c \right] (\bar{\lambda}^4)^r, \quad g(\bar{\lambda}) = \sum_{r=0}^{\infty} (-1)^r 4^{r+1} \left[ \frac{1}{(4r)!} + \frac{1}{(4r+3)!} c \right] (\bar{\lambda}^4)^r,$$

after replacements $n \to i \frac{\partial}{\partial \varphi}$ we come to differential with respect to $\varphi$ and difference with respect to $x$ non-local equation

$$L \Phi = 0, \quad L = (\nabla_{-2a} \nabla_{+2a} + 4) + G_\infty - F_\infty (\nabla_{-a} \nabla_{+a} + 2), \tag{5.8}$$

where the following notations for difference and differential operators are used:

$$\nabla_{\pm \xi} \Phi(x, \varphi) = \pm [\Phi(x \pm \xi, \varphi) - \Phi(x, \varphi)],$$

$$F_s = \sum_{r=0}^{s} \left[ \frac{4}{(4r)!} + \frac{-1}{(4r+3)!} L_c \right] (\bar{a}\Lambda)^{4r}, \quad G_s = \sum_{r=0}^{s} (-1)^r 4^{r+1} \left[ \frac{1}{(4r)!} + \frac{1}{(4r+3)!} L_c \right] (\bar{a}\Lambda)^{4r},$$

$$L_c = \frac{EJ\bar{a}^3}{B_x R^3} \frac{\partial^4}{\partial \varphi^4} \left( \frac{\partial^2}{\partial \varphi^2} + 1 \right)^2, \quad \Lambda^4 = -\frac{R}{B_x} \frac{\partial^4}{\partial \varphi^4} \left( \frac{\partial^2}{\partial \varphi^2} + 1 \right) \left[ p + \frac{D_\varphi}{R^3} \left( \frac{\partial^2}{\partial \varphi^2} + 1 \right) \right].$$

The single-field and the multi-field models based on Eq. (5.8) can be obtained in a way similar to that applied to the discrete systems, i.e., by using Taylor series expansion with respect to $x$, retaining up to fourth order derivatives (Vasiliev, 1993, 1994).

Comparison of the models is carried out for the solutions of the form $\Phi(x, \varphi) = \bar{\Phi} e^{i(kx+n\varphi)}$.

The equation of the single-field theory is the classical homogeneous equation for orthotropic shell with averaged characteristics when the stiffness of frame is smeared out along the cell of periodicity. This model is accurate for long-wave buckling modes with a half-wave containing several frames. The two-field model includes the equation of the single-field model and it has another equation that improves the single-field model giving an accurate approximation not only for small buckling mode wavenumbers $k$, but also for $k$ close to $\pi/a$. Accuracy of the models is illustrated by Fig. 17, where the dimensionless critical pressure $\bar{p} = p/p_a$, obtained by minimization of $\bar{p} = \bar{p}(n, k)$ on $n$, is presented as the function of $k$. The results obtained by using the structural model, the single-field, and the two-field models are marked by "0", "1" and "2", respectively.

## 6. Phase transitions: multi-field solitons

### 6.1. Elastically hinged molecule model

Elastically hinged molecule model (EHM model) was introduced by Dmitriev et al. (1997) and used for modeling domain walls in crystals (Dmitriev et al., 1997; Shigenari et al., 1997). The one-dimensional crystal is modeled by a chain of undeformable molecules (Fig. 18(a)) linked to each other by the elastic hinges with rigidity $c$. The chain is compressed by the axial force $p$. The transversal displacement of $n$ th hinge is denoted by $u_n$. Each hinge of the chain is in the external anharmonic potential $ku_n^2/2 + su_n^4/4$, $k > 0$, $s > 0$.

Equation of motion for $n$ th hinge has the form

$$m \frac{d^2 u_n}{dt^2} + c \frac{1}{h^2} (u_{n+2} - 4u_{n+1} + 6u_n - 4u_{n-1} + u_{n-2}) + p \frac{1}{h} (u_{n+1} - 2u_n + u_{n-1}) + ku_n + s u_n^3 = 0. \tag{6.1}$$

The role of the elastic hinges and the non-linear elastic supports is to keep the chain as a straight line while the compression force tends to destroy the horizontal arrangement of molecules. The competition between these two factors gives rise to modulation instability in the system.

### 6.2. N-periodic static structures

Obviously, equation of motion (6.1) has a trivial solution $u_n = 0$. A few types of $N$-periodic solutions, for which $u_n = u_{n+N}$, can be found.

The solution with the period $N = 2$ has the form

$$w_{2n} = -w_{2n-1} = Y_2, \quad Y_2 = \pm\sqrt{4P - 16F - 1}, \tag{6.2}$$

where $w_n = u_n\sqrt{s/k}$, $F = c/kh^2$, $P = p/kh$. The sign '+' corresponds to the structure with even nodes up and odd ones down and the sign '−' vice versa. This solution exists if $P > 4F + 1/4$.

The solution with the period $N = 3$ has the form

$$w_{3n} = Y_3, \quad w_{3n-1} = w_{3n-2} = \alpha Y_3, \quad Y_3 = \pm A, \quad A > 0. \tag{6.3}$$

This solution exists if $P > 3F + 1/3$. Again, we have '+' and '−' structures.

The '+' and '−' four-periodic solutions are

$$w_{4n} = w_{4n-1} = -w_{4n-2} = -w_{4n-3} = Y_4, \quad Y_4 = \pm\sqrt{2P - 4F - 1}, \tag{6.4}$$

which exists if $P > 2F + 1/2$.

*6.3. Multi-field soliton solutions*

Multi-field models can help to obtain approximate solutions describing moving domain walls in periodic structures. As an example, we will derive the domain wall solutions in the two-periodic structure.

Firstly, one can try to obtain solution using the single-field model. The single-field long-wave continuum analogue to the discrete equation (6.1) has the form

$$mu_{tt} + h^2\left(c + \frac{1}{12}ph\right)u_{xxxx} + phu_{xx} + k u + s u^3 = 0. \tag{6.5}$$

However, the single-field equation (6.5) does not support the static 2-periodic solution, Eq. (6.2).

We now use the equations written for the macrocell consisting of two hinges,

$$m\frac{d^2 u_n^{[1]}}{dt^2} + c\frac{1}{h^2}\left(u_{n+2}^{[1]} - 4u_{n+1}^{[2]} + 6u_n^{[1]} - 4u_{n-1}^{[2]} + u_{n-2}^{[1]}\right) + p\frac{1}{h}\left(u_{n+1}^{[2]} - 2u_n^{[1]} + u_{n-1}^{[2]}\right) + k u_n^{[1]} + s\left(u_n^{[1]}\right)^3 = 0,$$

$$m\frac{d^2 u_{n+1}^{[2]}}{dt^2} + c\frac{1}{h^2}\left(u_{n+3}^{[2]} - 4u_{n+2}^{[1]} + 6u_{n+1}^{[2]} - 4u_n^{[1]} + u_{n-1}^{[2]}\right) + p\frac{1}{h}\left(u_{n+2}^{[1]} - 2u_{n+1}^{[2]} + u_n^{[1]}\right) + k u_{n+1}^{[2]} + s\left(u_{n+1}^{[2]}\right)^3 = 0$$

and introduce the two continuum displacement functions, $u^{[1]}(x,t)$ and $u^{[2]}(x,t)$. Applying the procedure described above one obtains the two-field model,

$$mu_{tt}^{[1]} + Lu^{[1]} + \widetilde{L}(u^{[2]} - u^{[1]}) = 0,$$
$$mu_{tt}^{[2]} + Lu^{[2]} - \widetilde{L}(u^{[2]} - u^{[1]}) = 0, \tag{6.6}$$

where we have introduced the following operators:

$$Lu = h^2\left(c + \frac{1}{12}ph\right)u_{xxxx} + phu_{xx} + ku + su^3,$$

$$\widetilde{L}\Delta u = h^2\left(-\frac{1}{3}c + \frac{1}{12}ph\right)\Delta u_{xxxx} + (-4c + hp)\Delta u_{xx} + \frac{1}{h^2}(-8c + 2ph)\Delta u.$$

In this case, the equations of the two-field model cannot be uncoupled due to the presence of non-linearity.

Introducing dimensionless variables, $\tau = t\sqrt{k/m}$, $\xi = x/h$, and inserting $w^{[1]}(\xi,\tau) = -w^{[2]}(\xi,\tau) = -Y_2 = const$ into Eq. (6.6) one obtains the continuum analogue of the 2-periodic static solution for discrete system, Eq. (6.2).

Substituting

$$w^{[1]}(\xi,\tau) = -w^{[2]}(\xi,\tau) = W(\xi - v\tau) \tag{6.7}$$

in to Eq. (6.6), and keeping only second order derivatives, one obtains the equation $BW_{\eta\eta} + AW - W^3 = 0$, where $\eta = \xi - v\tau$, $A = 4P - 16F - 1$, $B = (P - 8F) - v^2$, which supports the following traveling kink solution:

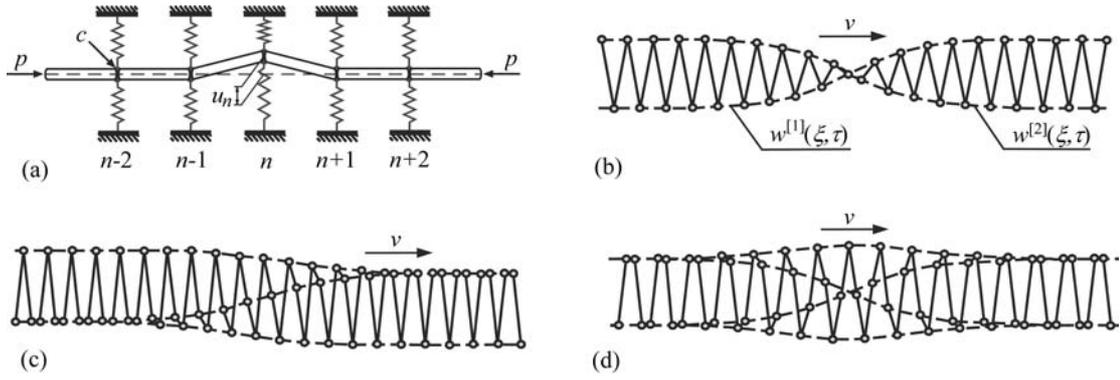

Fig. 18. (a) Chain of elastically hinged, rigid molecules in the nonlinear background potential. (b)-(d) Moving domain walls in the two-, three-, and four-periodic discrete structures (solid line), and their approximations by slowly varying interpenetrating field functions in the corresponding multi-field theories (dashed lines).

$$W(\eta) = \pm\sqrt{A}\tanh\left(\sqrt{\frac{AB}{2}}(\eta - \eta_0)\right). \tag{6.8}$$

This result can be rewritten in the form,

$$w_n(\tau) = \pm(-1)^n\sqrt{A}\tanh\left(\sqrt{\frac{AB}{2}}[(n - n_0) - v\tau]\right), \tag{6.9}$$

which gives an approximate, moving domain wall solution for the discrete system in dimensionless form. The solution describes a smooth conjugation between '+' and '−' structures and for $|n| \to \infty$ it gives the static 2-periodic solution (6.2). The solution exist for $A > 0$ and $B > 0$. The second condition defines the limiting value of the propagation velocity.

Figure 18(b) presents the domain wall solution in the 2-periodic structure with rapidly varying displacements of hinges in the discrete system (solid line) and the two-field approximation for this solution by two interpenetrating kinks (dashed lines).

The three- and four-field models describing the domain walls in the three- and in the four-periodic static solutions given by Eq. (6.3) and Eq. (6.4), respectively, were derived and studied in Dmitriev et al. (1997), Shigenari et al. (1997). Figures 18(c) and 18(d) show the moving domain wall solutions in the three- and four-periodic structures in discrete system defined by Eq. (6.1). The dashed lines show how the displacements of hinges are approximated by the interpenetrating slowly varying functions in the corresponding multi-field models.

## 7. Discussion and conclusions

We have considered the basic concepts, models, and methods of the multi-field continuum theory for the structured solids and periodic structures. Below we discuss possible applications, the reasons of the success and the directions of further development of the multi-field approach.

*7.1. Possible applications of the multi-field theory*

Generalized continuum models typically include the conventional continuum model as a particular case. However, the generalized modes are more complicated and their use is justified only in the cases when a simpler theory does not give an adequate result in describing some essential phenomena. One of the problems of the generalized mechanics is the search for the practical problems where the conventional theory does not work and to offer a theory capable of solving these problems.

Since the multi-field theory is valid for both long- and short-wavelength phenomena, i.e., for the phenomena occurring at both macro- and structural levels, it is a general theory that is capable to study any of them, and it is indispensable or even irreplaceable for the problems when the short and long waves are coupled.

Thus, the multi-field theories can be efficient for solving the problems with coupled long- and short-wavelength dynamics and the problems with coupled low- and high-frequency dynamics of bodies with microstructure. One of the attractive features of bodies with periodic structure is their capability to serve as filters for the elastic waves over certain frequency bands and polarizations. The need of optimization of structures for such purposes demands the development of models valid for low and high frequencies and long and short waves (Ruzzene and Scarpa, 2003, 2005; Gonella, 2007).

Multi-field models are useful in developing of continuum approximations for spatially localized short-wave, high frequency excitations that can exist and propagate in non-linear discrete systems. In physics they are known as intrinsic localized modes, or discrete breathers (Sievers and Takeno, 1988; Flach and Willis, 1998; Flach and Gorbach, 2008).

As a promising direction for application of generalized theories we note materials with complex microstructure that may result in their unusual properties. As an example, materials with negative Poisson ratio, or auxetics, were mentioned, and this field is now moving fast. Some results are summarized in the reviews (Konyok et al., 2004; Yang et al., 2004). Multiscale hybrid materials with negative Poisson's ratio have been discussed by Pasternak and Dyskin (2008). Unusual properties are often determined by the complex structure of a translational cell of the body and by a non-trivial response of the translational cell to external factors. Conventional models frequently cannot be used for modeling such behavior and the adequate generalized models should be used instead.

Fracture, instability, and plastic deformation often begin at structural level under relatively large stresses/deformations. Local deformations in the vicinity of boundaries, concentrated forces, defects or inhomogeneties can have monotonous or, in some cases, short wavelength character. In the latter case, the multi-field theory can be useful.

As another class of problems, let us note the problems of instability and phase transitions in structural bodies. Short-wavelength instabilities can often take place in such bodies. Compressive loading of thermal or mechanical origin may lead to appearance of short-wave displacements in structural solids. Buckling of internal or surface layers for sufficiently rigid background media may have a short-wavelength form because it may have energy smaller than the long-wave ones. The multi-field approach provides an adequate continuum modeling for the short-wavelength instability and for calculating the corresponding critical loads.

In Section 5 we have applied the multi-field theory to stability analysis of structural bodies with continuum translational cell. Multi-field dynamical models for layered media and layered sphere have been derived by Il'iushina (1972) and Molodtsov (1982). Let us note, that the idea of eliminating internal degrees of freedom is one of the interesting ideas for generalized and, in particular, non-local mechanics. This idea should be further developed in frame of the single-cell theories. Combination of this idea with macrocell method in the multi-field theory may be useful for deriving models that take into account the intra-cell, short- and long-wavelength deformations.

*7.2. Why does the multi-field approach works?*

From the physical point of view, the multi-field theory is based on the assumptions that change basic hypothesis of the conventional theory and hence one can expect new physical results from this theory.

Following discussion may help to understand why the multi-field models with low-order gradient terms provide a good approximation for both slowly and rapidly varying displacements in a discrete system. Long-wavelength displacements can be well described by a single slowly varying function, but the short-wavelength displacements can be described by a single function only if it varies rapidly. In the long-wavelength models only the low order gradient terms are used. Such models are correct for slowly varying displacements but they fail in describing rapidly varying ones. Figures 12, 15, and 18 demonstrate that both slowly and rapidly varying displacements can be accurately approximated by several slowly varying field functions. Such functions coincide when displacements in the discrete system vary slowly and split when they vary rapidly. Therefore, by increasing the number of fields, the multi-field approach gives a natural way to describe both long- and short-wavelength displacements by using slowly varying functions and study them in the framework of continuum mechanics.

Consideration of the dispersion surfaces of the discrete model and its continuum analogues gives another formal explanation of the success of the multi-field theory (Il'iushina, 1969; Vasiliev, 1993; Vasiliev et al., 2005). The $N$-field theory is constructed as a continuum analogue for the discrete system with a periodic cell contain-ing $N$ primitive translational cells. The number of equations in the discrete model increases by $N$ times but they do not contain any new information because the dispersion surfaces for discrete models in this case are just the original dispersion surfaces $N$ times folded in the $N$ times reduced first Brillouin zone. However, the $N$-field continuum model, derived for a periodic cell consisting of $N$ primitive cells, refines the single-field model because it gives a piecewise approximation of the exact dispersion surface. Each piece approximates the dispersion surface of the discrete system in the vicinity of the point $(K_x, K_y) = (0, 0)$ of the reduced first Brillouin zone. Thus, the $N$-field theory gives a good approximation of the dispersion surfaces not only for long waves but also for short waves and, in the case of multiple folding, for the waves inside the first Brillouin zone. This was illustrated in Section 2 for the simple chain and in Section 3 for the Cosserat lattice.

*7.3. Notes on further development of the multi-field theory*

We have discussed some achievements of the multi-field theory. Derivation and analysis of generalized continuum models on the basis of structural models is one of the approaches to develop phenomenological theories and to find their interpretations and applications. This is the reason why in our works we have focused on the structural physical models.

The multi-field theory can follow the directions of development of the other field theories. In Section 1.1. of the Introduction some of those directions have been already mentioned: development of the theory on the bases of a structural model or phenomenologically, development of analytical and numerical methods of analysis, development of experimental methods, applications of the theory. These branches of the field theory have been developed to a large extent for micropolar, non-local and higher-order gradient theories. Experimental methods for these theories have been described, e.g., by Lakes (1995).

Historically, most of the branches of solid mechanics have been developed before the generalized continuum theories were developed. After the micropolar, the non-local, and the higher-order gradient theories were developed, they have influenced considerably such branches of solid mechanics as fracture mechanics and plasticity theory, and they have found numerous applications in modeling of structural solids taking into account the microstructure effects (see, for example, articles cited in Section 1.2 of Introduction). Similarly, some corrections of conventional solid mechanics may be considered in cases when the essential effects, captured by the multi-field theory, take place.

Overall, changing the basic hypothesis of the conventional theory should lead to interesting new features of the generalized theories and they should be studied and discussed. This was one of the leading ideas for the authors in their efforts.

**Acknowledgments**

SVD gratefully acknowledges a financial support provided by the DST-RFBR joint grant 08-02-91316-Ind-a and the RFBR grant 09-08-00696-a.